\documentclass[ste,twoside]{stefano}

\usepackage{amsmath}
\usepackage{amssymb}
\usepackage{graphics}
\usepackage{rotating}
\usepackage{cite}
\usepackage{color}

\def\makeheadbox{{%
\hbox to0pt{\vbox{\baselineskip=10dd\hrule\hbox
to\hsize{\vrule\kern3pt\vbox{\kern3pt \hbox{  {\sc hep-ph/0208086}
} \hbox{ {\sc Int. J. Mod. Phys. A {\bf 19}, 677-694 (2004)}
\hspace*{7.0cm} {\color{blue}{$\boldsymbol{\Sigma \delta
\Lambda}$}} }
\kern3pt}\hfil\kern3pt\vrule}\hrule}%
\hss}}}

\def\0{\mbox{\tiny $0$}}
\def\1{\mbox{\tiny $1$}}
\def\2{\mbox{\tiny $2$}}
\def\3{\mbox{\tiny $3$}}
\def\4{\mbox{\tiny $4$}}
\def\5{\mbox{\tiny $5$}}
\def\6{\mbox{\tiny $6$}}
\def\7{\mbox{\tiny $7$}}
\def\8{\mbox{\tiny $8$}}
\def\9{\mbox{\tiny $9$}}
\def\a{\mbox{\tiny $\alpha$}}
\def\b{\mbox{\tiny $\beta$}}
\def\c{\mbox{\tiny $c$}}
\def\st{\mbox{\tiny $st$}}
\def\in{\mbox{\tiny $in$}}

\def\osc{\mbox{\tiny $osc$}}
\def\n{\mbox{\tiny $n$}}
\def\e{\mbox{\tiny $e$}}
\def\fp{\mbox{\tiny $f$}}
\def\foh{\mbox{\tiny $\frac{1}{2}$}}
\def\f14{\mbox{\tiny $\frac{1}{4}$}}
\def\infm{\mbox{\tiny $-\infty$}}
\def\infp{\mbox{\tiny $+\infty$}}
\def\mi{\mbox{\tiny $-$}}

\begin{document}
%

\title{WAVE PACKETS AND QUANTUM OSCILLATIONS}

\author{
Stefano De Leo\inst{1}
\and
Celso C. Nishi\inst{2}
\and
Pietro P. Rotelli\inst{3}
}

\institute{
Department of Applied Mathematics, State University of Campinas\\
PO Box 6065, SP 13083-970, Campinas, Brazil\\
{\em deleo@ime.unicamp.br}
\and Department of Cosmic Rays and Chronology, State University of
Campinas\\
PO Box 6165, SP 13083-970, Campinas, Brazil\\
{\em ccnishi@ifi.unicamp.br}
\and
Department of Physics, INFN, University of Lecce\\
PO Box 193, 73100, Lecce, Italy\\
{\em rotelli@le.infn.it}
}


\date{{\em August, 2003}}

\abstract{ We give a detailed analysis  of the oscillation formula
within the context of the wave packet formalism. Particular
attention is made to insure flavor eigenstate creation in the
physical cases ($\Delta p \neq 0$). This requirement imposes non
instantaneous particle creation in all frames. It is shown that
the standard formula is not only exact when the mass wave packets
have the same velocity, but it is a good approximation when
minimal slippage occurs. For more general situations the
oscillation formula contains additional arbitrary parameters,
which allows for the unknown form of the wave packet envelope. }









\maketitle


\section*{I. INTRODUCTION}

In the last few years, the great interest in neutrino physics, and
in particular in neutrino masses, has stirred  up an increasing
number of theoretical works upon the quantum mechanics of
oscillation phenomena~\cite{ZRA98,ZUB98,BCG99}. Notwithstanding
the exceptional ferment in this field, the situation is still
confused, and the conceptual difficulties hidden in the
oscillation formulas represent an intriguing, and sometimes
embarrassing, challenge for physicists. The most controversial
point in discussing the quantum mechanics of particle oscillations
is represented by the derivation  of formulas containing {\em
extra factors} in the  oscillation
length~\cite{LIP95,GL97,TTTY99,OT00,DDR00,TTTY00,FIE01,GIU02,ROT00}.
In this paper, we review the source of these factors and show why
in the wave packet formalism with {\em minimal slippage} one
finds, independently from the kinematical assumptions, the
standard oscillation probability. However, in more general
situations one does indeed find different oscillation formulas.
This question is of course essential if one wants to derive
consistent mass differences from oscillation phenomena. It may
well be that different experiments, or even sets of measurements
within a given experiment (e.g. atmospheric neutrino data),
involve different oscillation formulas, and this must be taken
into account.

 One of the basic assumptions in neutrino physics is that only flavor
eigenstates are destroyed or created. Now, in the wave function
formalism this is a problem which, in our opinion, has not yet
been satisfactorily solved. The most common approach
is to assume instantaneous creation within
the context of an equal momentum hypothesis~\cite{KAY00}.
Unfortunately, there is no physical Lorentz frame in which this
occurs. Some authors have even tried to bypass the problem by {\em
re-defining} the flavor eigenstates according to
convenience~\cite{GIUPRD}. In this paper we shall describe how to
achieve this obligatory condition within the wave packet
formalism. It will automatically imply {\em non instantaneous}
creation of the wave packet for any physical production process.

In our analysis, we shall for simplicity work within a two
flavor/mass mixing model.
The structure of the paper is as follows:\\
$\cdot$  In Section II, we recall the arguments in favor of extra
factors in the oscillation formula. We shall in particular note
that an essential assumption in these derivations is that the
flavor eigenstate is {\em identical}, including
  the phase, at each point of creation.\\
$\cdot$ In Section III, we recall the
equal momentum wave packet derivation of the standard oscillation
formula.\\
$\cdot$ We discuss the physical kinematics for particle creation in
Section IV.\\
$\cdot$  The extension from the equal momentum case to an arbitrary
case is tackled, and solved, in Section V. In this Section, we
also show that, while physically essential, this generalization does
not in itself invalidate the standard oscillation formula.\\
$\cdot$ In Section VI, we return to the question of non-standard oscillation
formulas by showing the results of a multiple peak (specifically a
two-peak) model with substantial slippage.\\
$\cdot$ We draw our conclusions
in Section VII.

\section*{II. EXTRA FACTORS IN THE OSCILLATION FORMULAS}

In the quantum mechanics of particle oscillation, substantial
mathematical simplification results from the assumption that the
space dependence of the wave functions is one-dimensional, hence,
in what follows, we shall use this simplification. Flavor
oscillations are observed when a source creates a particle which
is a mixture of two or more mass eigenstates. The main aspects of
oscillation phenomena can be understood by studying the two flavor
problem
\[
\left( \begin{array}{c}   \boldsymbol{\nu_{\a}} \\
                          \boldsymbol{\nu_{\b}}
\end{array} \right) =
 \left( \begin{array}{rr}   \cos \theta &  \, \sin \theta \\
                          $-$ \sin \theta & \, \cos \theta
\end{array} \right) \,
\left( \begin{array}{c}   \boldsymbol{\nu_{\1}} \\
                          \boldsymbol{\nu_{\2}}
\end{array} \right)~,
\]
where $\boldsymbol{\nu_{\a}}$ and $\boldsymbol{\nu_{\b}}$ are
flavor eigenstates and $\boldsymbol{\nu_{\1}}$ and
$\boldsymbol{\nu_{\2}}$ are mass eigenstates.

Suppose we have a physical system whose initial state is represented
by the  flavor state  $\boldsymbol{\nu_{\a}}$.
At later times, the  probability to find  the flavor state
$\boldsymbol{\nu_{\b}}$ is conventionally expressed in terms of the mixing
angle
$\theta$ and of the relative phase  $\Delta \Phi$ by
\[
P \left( \boldsymbol{\nu_{\a}}  \to \boldsymbol{\nu_{\b}} ; \, t
\right) =
\sin^{\2} 2 \theta \, \, \frac{ 1 -
 \, \cos \Delta \Phi}{2}~.
\]
It is to be noted that the above formula ignores any possible
effects involving the shape of the wave functions. The Lorentz
invariant phase factor $\Delta \Phi$ is usually given in terms of
the distance $L$ (travelled in the time $T$), of the mass
difference $\Delta m^{\2}$, and of the energies mean value
$\bar{E}$. Due to the relativistic nature of neutrinos, this phase
is evaluated by considering $T \approx L$ and $p_{\1,\2} \approx
E_{\1,\2}$, i.e.
\begin{equation}
\label{sf}
\Delta \Phi =   T \,
\Delta E - L \, \Delta p
\approx
L \, \left( \Delta E - \Delta p \right)
\approx
L \, \Delta m^{\2} / 2 \bar{E}~.
\end{equation}
By using this phase difference, one gets the well-known
expression~\cite{BPo78,BPe87,KGDP89,BV87,KG93}
\begin{equation}
\label{po}
P \left( \boldsymbol{\nu_{\a}}  \to \boldsymbol{\nu_{\b}} ; \,  L
\right) \approx
\sin^{\2} 2 \theta \, \sin^{\2} \left( \frac{\Delta m^{\2}}{4 \bar{E}} \, L
\right) = \sin^{\2} 2 \theta \, \sin^{\2}
\left( \pi \, \frac{L}{L_{\osc}} \right)~,
\end{equation}
where
\begin{equation}
\label{losc} L_{\osc} = \frac{4 \pi \bar{E}}{\Delta m^{\2}}
\approx 2.48 \, \frac{E [\mbox{GeV}]}{ \Delta m^{\2}
[\mbox{eV}^{\2}]} \, \mbox{Km}
\end{equation}
is the distance at which $\Delta \Phi$ becomes $2 \pi$. In this
paper, we shall refer to Eq.(\ref{po}) as the {\em standard
oscillation probability} and to Eq.(\ref{losc}) as the {\em
standard  oscillation length}.  As an aside, we note that the
assumptions in Eq.(\ref{sf})  have formally cost us the Lorentz
invariance of the standard oscillation formula. As written, it is
no longer valid in all Lorentz frames.

The historical development of the calculation of the
phase factor $\Delta \Phi$ is interesting and a little mysterious.
The
first prediction of the oscillation probability was  given in 1969
by Gribov and Pontecorvo~\cite{GP69}
\[
|\nu_{\e}(t)|^{\2} =  |\nu_{\e}(0)|^{\2} \, \left[ \left( 1 -
\frac{\sin^{\2} 2\theta}{2} \right) + \frac{\sin^{\2} 2\theta}{2}
\, \cos \left( \boldsymbol{2} \, \frac{\Delta m^{\2}}{2 \,
\bar{p}} \, t \right) \right]~.
\]
The important point to note here is a factor two difference
with respect to the relative phase which appears in Eq.(\ref{sf}).
Only some years later,
to be more  precise in 1976, was the standard oscillation probability
obtained by Fritsch and  Minkowsky~\cite{FM76}.

For a long time thereafter, the oscillation probability (\ref{po})
stood as the fundamental starting point in neutrino physics and no
comment was ever made on the additional factor two which appeared
in the Gribov and  Pontecorvo work (for further details see
ref.~\cite{FIE01}).  In 1995, Lipkin re-discussed the derivation
of the oscillation probability and pointed out that, by assuming
the {\em equal momentum scenario}, an extra factor of two appears
in the oscillation phase~\cite{LIP95}. The hypothesis of equal
energies was then suggested to re-obtain and justify the standard
result~\cite{GL97}. However, simply by following the reasoning of
Lipkin~\cite{LIP95}, the authors of ref.~\cite{DDR00} showed that,
contrary to Lipkin's assertion, the only case in which the
standard oscillation phase can be reproduced is in the {\em equal
velocity scenario}. This condition also distinguishes itself from
the others (equal energy or equal momentum) by being Lorentz
invariant. This does not mean that this scenario, which yields the
maximum simplicity and recovers the standard oscillation
probability, coincides with the real situation in neutrino
production. As observed in refs.~\cite{OT00,DDR00} common
velocities imply $E_{\1}/E_{\2} = p_{\1}/p_{\2} = m_{\1} / m_{\2}$
and since this may be very far from unity it would contradict the
estimates of the neutrino energies for the known production
mechanisms. A discussion of the kinematical constraints derived
from energy-momentum conservation in neutrino production will be
given in Section IV.

In order to understand how extra factors appear in the oscillation
formula, let us analyze the difference of phase responsible for
the oscillation phenomenon. In the plane wave formalism the
appropriate plane wave phase is associated with each mass
eigenstate. Since the four-momentum of different masses cannot
coincide, the phase of each mass eigenstate will change with time
and distance. Thus an initially pure flavor-eigenstate will be
modified with time. The mass-eigenstate phase difference is
\begin{equation}
\label{int}
\Delta \Phi  = \Delta \left( E \, T - p \, L  \right)~.
\end{equation}
In the standard treatment, one evaluates this by setting $\Delta T
= \Delta L = 0$~\cite{LBBRSG96,KW98}. Nevertheless, if the two
mass eigenstates have {\em different} speeds, by assuming
instantaneous creation,  we should experience, at the common time
$T$, the interference between wave  function points which have
travelled different distances, i.e.
\[
L_{\1}=v_{\1} \, T~~~\mbox{and}~~~L_{\2}=v_{\2} \, T~~~~~\left[~\Rightarrow~~
\Delta L = T \, \Delta v~~~\mbox{and}~~~T =  \bar{L} / \bar{v}~\right]~.
\]
The interference between wave function points which have travelled
different distances is the source of an extra multiplicative
factor,
\begin{equation}
\label{ef} \epsilon = \frac{\Delta \Phi ( \Delta L \neq 0
)}{\Delta \Phi ( \Delta L = 0 )} ~~~~~ \left[~\equiv \frac{\Delta
\Phi ( \Delta v \neq 0 )}{\Delta \Phi ( \Delta v = 0 )}~ \right]~,
\end{equation}
in the oscillation phase (length). In order to quantify this
effect, we explicitly calculate the difference of phase given in
Eq.(\ref{int}). By simple algebraic manipulations, we obtain
\begin{eqnarray}
\label{intl}
\Delta \Phi & = &
\left( \, \frac{1}{\bar{v}} \, \Delta E - \Delta p -
\frac{\bar{p}}{\bar{v}} \, \Delta v \,  \right) \,
 \bar{L}~.
\end{eqnarray}
As particular cases, we can immediately compute the Lorentz invariant factor
$\Delta \Phi$ in
the common velocity and common energy scenario,
\begin{eqnarray*}
\Delta \Phi \left[ \, \, \Delta v  =  0 \, \,  \right]   & = &
\frac{\Delta m^{\2}}{2 \bar{p}} \, \bar{L}
\approx
\frac{\Delta m^{\2}}{2 \bar{E}} \, \bar{L}
~~~~~\left[ \, \mbox{standard result}
\, \right]~,\\
\Delta \Phi \left[ \,  \Delta E = 0  \,  \right]
& = &
\frac{\Delta m^{\2}}{\bar{p}} \, \bar{L}
\approx
\frac{\Delta m^{\2}}{\bar{E}} \, \bar{L}
~~~~~\left[ \,
\mbox{extra factor two} \, \right]~.
\end{eqnarray*}
In order to get a more general expression for the extra
factor, we rewrite $\epsilon$  in terms of $\bar{p}$, $\bar{E}$,
$\Delta p$ and $\Delta E$ as follows
\begin{equation}
\label{a}
\epsilon \left( \, \bar{p} \, , \, \bar{E} \, ; \, \Delta p \, , \, \Delta E
\, \right)
=  1 + \frac{\bar{p} \Delta v}{\bar{v} \Delta p - \Delta E}~.
\end{equation}
Considering $\Delta p \ll \bar{E}$ and $\Delta E \ll \bar{E}$, we obtain
\begin{eqnarray}
\label{a2}
\epsilon
\left( \, \bar{p} \, , \, \bar{E} \, ; \, \Delta p \ll \bar{E} \, ,
\, \Delta E \ll \bar{E} \right)
& ~=~ & 1 +
\frac{\, \bar{p} \left(
\bar{E} \Delta p - \bar{p} \Delta E  \right) }{\bar{E} \left(
\bar{p} \Delta p -
\bar{E} \Delta E \right)} \,
\left\{ 1 + \mbox{O} \left[ \frac{\Delta E \Delta p}{\bar{E}^{\2}} \, ,
\,  \left( \frac{\Delta E}{\bar{E}} \right)^{\2} \right] \right\} \nonumber
\\
 & \approx & 1 +
\frac{\, \bar{p} \left(
\bar{E} \Delta p - \bar{p} \Delta E  \right) }{\bar{E} \left(
\bar{p} \Delta p -
\bar{E} \Delta E \right)} \\
 & \approx & \left\{ \begin{array}{ll}
1 ~~~\mbox{when}~~\Delta p = 0
~~\mbox{and}~~p_{\1,\2} \ll E_{\1,\2}~,\\
 ~\\
2 ~~~\mbox{when}~~\Delta p = 0~~\mbox{and}~~p_{\1,\2} \approx
E_{\1,\2}~.
\end{array} \right.
\nonumber
\end{eqnarray}
There are difference in the value of $\epsilon$ between the
scenarios of common momentum and common energy but they are only
significant in the non-relativistic limit. Such a situation could
in principle be tested, for example, in the neutral kaon system.
In neutrino physics, where non-relativistic neutrinos are
unobservable, when $\Delta v \neq 0$, we practically {\em always}
find an extra factor two in the oscillation length. In conclusion,
to recover the standard formula for neutrinos in this formalism
(where the energy of production is approximately known and is
orders of magnitude greater than the postulated masses), in
addition to the exact common velocity scenario we would also need
to impose almost equal masses to guarantee $E_{\1}/ E_{\2} \approx
1$~\cite{DDR00}.

Now it is important to observe that, in all the above, one has
implicitly assumed that the flavor eigenstate is always given by
the mixing matrix (chosen real by convention) with which we
started this Section. That is, the flavor eigenstate has been
assumed {\em identical} at all points and/or times of creation.
However, we can of course multiply a flavor (mass) eigenstate by a
phase factor without modifying its flavor (mass). Perhaps less
obvious, this phase may even be space-time dependent. A
significant example of this occurs in the next Section and is
generalized in Section V. An alternative wave packet example,
specifically devised to approximate at a given time the above non
standard oscillating phase, is presented in Section VI.

\section*{III. WAVE PACKET FORMALISM WITH INSTANTANEOUS CREATION}
In the preceding section, we introduced the fundamental arguments
leading to extra factors in the oscillation probability. In this
section, we are going to show why these factors do not appear in
the usual wave packet formalism. We begin by trying to understand
qualitatively the problem in a very simple case, that is $\Delta p
=0$~\cite{KAY81,RIC93}, and deduce from it several important
conclusions.

So far, we have only  considered a single plane wave. Rigorously,
such an energy-momentum eigenstate cannot represent a physical
state - it is not normalizable. It would also pose us with the
problem of defining $L$ and $T$ in the oscillation phase.
 To avoid these problems, we must employ a normalized
 superposition of plane waves
\[
\exp \left\{ -i \left[ E(p,m_{\n})t - p x
\right]  \right\}
\]
and describe the time evolution of flavor states by  the wave packet
\begin{eqnarray}
\label{psi}
\boldsymbol{\Psi}(x,t) & =  & \psi_{\1}(x,t) \,
\cos \theta ~ \boldsymbol{\nu_{\1}} +
\psi_{\2}(x,t) \, \sin \theta ~\boldsymbol{\nu_{\2}} \nonumber \\
 &  =  &
\left[ \, \psi_{\1}(x,t) \, \cos^{\2} \theta + \psi_{\2} (x,t) \,
\sin ^{\2} \theta
\, \right] ~\boldsymbol{\nu_{\a}} +
\left[ \, \psi_{\2}(x,t) -  \psi_{\1}(x,t) \, \right]
\cos \theta \, \sin \theta ~ \boldsymbol{\nu_{\b}} \nonumber \\
 & = & \psi_{\a}(x,t;\theta)~\boldsymbol{\nu_{\a}} +
\psi_{\b}(x,t;\theta)~\boldsymbol{\nu_{\b}}~,
\end{eqnarray}
where
\[
\psi_{\n} (x, t) =   \mbox{$\frac{1}{\sqrt{2 \pi}}$} \,
\int_{\infm}^{\infp} \hspace*{-0.3cm}
\varphi_{\n}(p ) \,
\exp \left\{ -i \left[ E(p,m_{\n})t - p x
\right] \right\} \, \mbox{d} p~.
\]
As a model assumption, we suppose that the momentum distributions
$\varphi_{\n}(p)$  are given by Gaussian functions peaked around
the mass eigenstate momenta $p_{\n}$, i.e.
\begin{equation}
\varphi_{\n}(p) =
 \left( \frac{ \, a^{\2}_{\n}}{2 \pi } \right)^{\f14} \,
\exp \left[ - a^{\2}_{\n}
\left( p - p_{\n} \right)^{2} / 4 \right]~.
\end{equation}
To {\em instantaneously} create at $t=0$, in a localized region
centered around the spatial coordinate $x=0$, a flavor state
$\boldsymbol{\nu_{\a}}$, we have to impose the following
constraint
\begin{equation}
\label{c100}
\psi_{\1}(x,0) = \psi_{\2}(x,0)~~~[~\Rightarrow~\Delta a = \Delta p =0~]~.
\end{equation}
Consequently, from Eq.(\ref{psi}) we get
\begin{equation}
\label{psi0}
\boldsymbol{\Psi}(x,0) =
\left( \frac{2}{\, \pi a^{\2} } \right)^{\f14}
 \,
\exp \left[ - \frac{\, \, x^{\2}}{\, \, a^{\2}} \right] \,
\exp \left[ i \, p_{\0}  x \right] \, \boldsymbol{\nu_{\a}}
= \psi(x,0) \, \boldsymbol{\nu_{\a}}~,
\end{equation}
where $p_{\0}=p_{\1}=p_{\2}$. The probability $P
(\boldsymbol{\nu_{\a}}  \to \boldsymbol{\nu_{\b}}  ;  \, t )$ of
observing a flavor state $\boldsymbol{\nu_{\b}}$  at the instant
$t$ is equal to the integrated squared  modulus of the
$\boldsymbol{\nu_{\b}}$  coefficient in Eq.(\ref{psi}),
\begin{eqnarray}
\label{sf2} P \left( \boldsymbol{\nu_{\a}}  \to
\boldsymbol{\nu_{\b}}  ; \, t \right) & = & \int_{\infm}^{\infp}
\hspace*{-0.3cm}
 | \psi_{\b} (x, t;\theta)|^{\2} \, \mbox{d} x~.
\nonumber \\
 & = & \sin^{\2}
2 \theta \, \, \left[ \, 1 - \mbox{Re} \int_{\infm}^{\infp}
\hspace*{-0.3cm}
 \psi_{\1} (x, t) \,
 \psi_{\2}^{*} ( x, t) \, \mbox{d} x  \, \right] / 2 ~.
\end{eqnarray}
Actually, this result with a unique time assumes that the detector
is not localized in a region smaller or comparable to the size of
the wave packet. Otherwise for $t$ we would have to use the
average time of measurement. In order to calculate the oscillation
probability, let us change the $x$-integration into
$p$-integration
\[
 \int_{\infm}^{\infp} \hspace*{-0.3cm}
 \psi_{\1} (x, t) \,
 \psi_{\2}^{*} ( x, t) \, \mbox{d} x =
\int_{\infm}^{\infp} \hspace*{-0.3cm}
\varphi^{\2} (p ) \,
\exp \left\{ - i \left[ E(p,m_{\1}) - E(p,m_{\2}) \right] t \right\} \,
\mbox{d} p
\]
and use the following approximation
\begin{eqnarray*}
E(p,m_{\1}) - E(p,m_{\2}) & = &
\left( 1 + \frac{p^{\2} - p_{\0}^{\2}}{E_{\1}^{^{\2}}} \right)^{\foh}
E_{\1}  -
\left( 1 + \frac{p^{\2} - p_{\0}^{\2}}{E_{\2}^{^{\2}}} \right)^{\foh}
E_{\2}\\
&  \approx & \left( 1 - \frac{p^{\2} - p_{\0}^{\2}}{2 \,
\bar{E}^{^{\2}}} \right) \, \Delta E \\
& \approx & \left[ 1 - \frac{\left( p - p_{\0} \right) \,
p_{\0}}{\bar{E}^{^{\2}}} \right] \, \Delta E\\
 & \approx & \Delta E + (p - p_{\0}) \Delta v~.
\end{eqnarray*}
This approximation is justified if we assume
$\delta p  \ll \bar{E}$ and  $\Delta E \ll \bar{E}$.
The oscillation term is then given by
\begin{eqnarray*}
 \int_{\infm}^{\infp} \hspace*{-0.3cm}
 \psi_{\1}^{*} (x, t) \,
 \psi_{\2} ( x, t) \, \mbox{d} x
 & \approx &
\left( \frac{ \, a^{\2}}{2 \pi } \right)^{\foh} \, \exp \left[ - i
\, \Delta E \, t \right] \, \int_{\infm}^{\infp} \hspace*{-0.3cm}
\exp \left[ - a^{\2} \sigma^{\2} / 2 \right] \, \exp \left[ - i \,
\sigma \,  \Delta v \,
t \right] \, \mbox{d} \sigma\\
& =  & \exp \left[ - i \, \Delta E \, t \right] \, \exp \left[ -
\,  \left( \, \frac{\Delta v \, t}{a \sqrt{2}} \right)^{\2}
\right]~.
\end{eqnarray*}
By observing that $\Delta E = \Delta m^{\2}/2 \bar{E}$ and using
the approximation  $T=\bar{L}/\bar{v}\approx \bar{L}$ (where $T$
stands for the observation time), Eq.(\ref{sf2}) reduces to
\begin{equation}
\label{sf2bis} P \left( \boldsymbol{\nu_{\a}}  \to
\boldsymbol{\nu_{\b}} ; \, \bar{L} \right) \approx \sin^{\2} 2
\theta \, \left\{ 1 - \exp \left[ - \,  \left( \, \frac{\Delta v
\, T}{a \sqrt{2}} \right)^{\2} \right] \, \cos \left( \frac{\Delta
m^{\2}}{2 \bar{E}} \, \bar{L} \right) \right\} \, / \, 2~.
\end{equation}
Thus, when minimal slippage occurs ($\Delta v \, T \ll a $) the
standard oscillation probability (\ref{po}) is a good
approximation and {\em no} extra factor appears in the oscillation
term . This contradicts the result given in the previous Section,
where, in the equal momentum scenario by using the  plane wave
derivation,  an extra factor of two was obtained, see
Eq.(\ref{a2}). To explain this apparent paradox, we observe that,
at time $T$ and at a fixed position $x_{\fp}$  in the overlapping
region,  we experience the interference between space points whose
separation at creation is given (see Fig.\,1) by
\[ \Delta  x_{\in} = - \Delta v \, T~.\]
This implies that an {\em additional} initial phase,
\begin{equation}
\label{idf}
\Delta \Phi_{\in} = - p_{\0} \, \Delta  x_{\in} =
 p_{\0} \, \Delta v \, T~,
\end{equation}
is automatically included in
the wave packet formalism. Consequently, the final
result contains both the phase difference
calculated in the previous Section, i.e.
\[
\Delta \Phi  = T \, \Delta  E   - p_{\0} \,  \Delta L~,
\]
and the additional term given in Eq.(\ref{idf}). Thus, the {\em
standard} result,
\begin{equation}
\Delta \Phi_{\st}  = \Delta \Phi_{\in} + \Delta \Phi = T \, \Delta
E~,
\end{equation}
is obtained. Hence, the difference in this scenario with that of
the previous Section is that here the flavor eigenstate is {\em
not} unique at all points of creation. {\em Each point is
associated with an appropriate $x$-dependent phase}.

Before proceeding further, we must ask whether the above  equal
momentum scenario is physically possible, and if not, how it is to
be modified while maintaining the creation of only a flavor
eigenstate. These modification could well change the oscillation
phase. To respond to these questions we must first review the
kinematics of particle creation.

\section*{IV. KINEMATIC CONSTRAINTS IN PRODUCTION}

We start by observing that any production process of a particle
(be it an oscillating particle or not) can be considered, for
kinematic purposes,  as {\em an effective two-body decay} such as
 \[ m_{\0}  \longleftarrow
 \overset{M}{\bullet} \longrightarrow m_{\1,\2}\]
where we recall that the subscript in $m_{\1,\2}$ refers to the
oscillating mass eigenvalues. If the production process is a decay
into more than two particles,  then $m_{\0}$ represents the
effective mass of all the accompanying particles and is of course
greater than or equal to the sum of their masses. For production
processes other than decays $M$ is simply the center of mass
energy and not the mass of a resonance. In this ``rest'' frame,
energy and momentum conservation imply
\[ M = \sqrt{p_{\n}^{\2} + m_{\0}^{\2}} +
\sqrt{p_{\n}^{\2} + m_{\n}^{\2}} = \sqrt{p_{\n}^{\2} + m_{\0}^{\2}} + E_{\n}
~,~~~~~\mbox{\small $n=1,2$}~.
\]
By simple algebraic manipulations,
we find~\cite{WIN81,GIU01}
\begin{equation}
\label{tbd}
\Delta E = \frac{\Delta m^{\2}}{2 M}~,~~~
\Delta p =\frac{\bar{E} - M}{\bar{p}} \,  \Delta E~~~\mbox{and}~~~
\Delta v = \frac{1  - (\bar{p} / \bar{E})^{^{\2}}
- (M / \bar{E})}{\bar{p}}\,
\left[ 1  - \left( \frac{\Delta E}{2 \bar{E}} \right)^{\2} \,
\right]^{\mi \1}
\, \Delta E~.
\end{equation}
The next  step is to observ that by assumption
\[ \Delta m \neq 0~~~\mbox{and thus}~~~
2 ( M - \bar{E} ) = \sqrt{p_{\1}^{\2}+m_{\0}^{\2}} +
\sqrt{p_{\2}^{\2}+m_{\0}^{\2}} \neq 0~.\]
Consequently, in the
rest frame of the decaying particle of mass $M$ (or, in general,
in the center of mass frame) we have
\begin{equation}
 \Delta E \neq 0~,~~~\Delta p \neq 0~~~\mbox{and}~~~\Delta v \neq 0~.
\end{equation}
This implies that there does not exist any frame in which $\Delta
v=0$ since this is a Lorentz invariant condition. We can also show
that there does not exist any frame in which $\Delta p=0$. In
fact, by performing a Lorentz transformation with velocity $\beta$
from the rest frame of the decaying particle $M$ (or center of
mass), we find
\begin{eqnarray*}
\Delta E' & = & \gamma \left( \Delta E - \beta \Delta p \right)~,\\
\Delta p' & = & \gamma \left( \Delta p - \beta \Delta E \right)~.
\end{eqnarray*}
To satisfy $\Delta p'=0$, we have to impose the following {\em unphysical}
condition on $\beta$
\[
|\beta| = |\Delta p / \Delta E| =  |( \bar{E} - M ) / \bar{p}| =
\left( \sqrt{p_{\1}^{\2}+m_{\0}^{\2}} + \sqrt{p_{\2}^{\2} +
m_{\0}^{\2}} \right) / 2 \bar{p} \,
> 1~.
\]
This shows that $(\Delta E , \Delta p )$ is  space-like.
Therefore, there will, on the contrary, always be frames in which
$\Delta E'= 0$.

It is now important to realize that the condition $\Delta p \neq
0$ automatically implies for $x\neq 0$ that the mass eigenstates
defined in the previous Section are no-longer equal at $t=0$,
\[ \psi_{\1}(x,0) \neq \psi_{\2}(x,0)~.\]
This inevitably leads, within the context of instantaneous
creation, to a non zero probability to find (see Fig.\,2) a flavor state
$\boldsymbol{\nu_{\b}}$ at time $t=0$~\cite{TTTY00,KG93}.
Indeed, for
 $\Delta p \neq 0$ and with instantaneous creation,
we  obtain the following oscillation probability
\begin{eqnarray}
\label{nzt}
P \left( \boldsymbol{\nu_{\a}}  \to
\boldsymbol{\nu_{\b}} ; \, t \right) &
 \approx  &
\sin^{\2} 2 \theta \, \left\{ 1 - \exp \left[ - \, \left( \frac{a
\, \Delta p}{2 \sqrt{2}} \right)^{\2} - \, \left( \, \frac{\Delta
v \, t}{a \sqrt{2}} \right)^{\2} \right] \, \cos \left(
\frac{\Delta m^{\2}}{2 \bar{E}} \, t \right) \right\} \, / \, 2~.
\end{eqnarray}
Thus, there does not exist {\em any} time for which the state is a
pure flavor eigenstate. In the next Section, we shall describe
how by generalizing to
 non instantaneous creation we can eliminate the initial difference
of phase between the mass eigenstates  and hence achieve
 {\em pure flavor creation event-wise }.

\section*{V. NON INSTANTANEOUS CREATION}
We have identified the initial difference of phases in the mass
eigenstates
\[
- \Delta p\, x_{\c}~,
\]
where $x_{\c}$ is a generic space point in the creation wave
packet, as the cause for having at the time of creation a state
which is {\em not} a pure flavor eigenstate. This undesired effect
can be removed either by the unphysical assumption of equal
momenta or by introducing for each space point $x_{\c}$ a
corresponding creation time $t_{\c}$ which satisfies the following
relation
\begin{equation}
\Delta E\, t_{\c} -
 \Delta p\, x_{\c}=0~.
\end{equation}
This condition guarantees
\[
dP \left( \boldsymbol{\nu_{\a}}  \to \boldsymbol{\nu_{\b}} ; \, x_{\c}, \,
t_{\c} \right) = 0
\]
and consequently allows for pure flavor creation {\em event-wise}.

Somewhat surprisingly this substantial modification does not
invalidate the standard oscillation formula. In fact, by following
the plane wave phase calculation of Section II, we find at the
interference space-time point $(x_{\fp},T)$ the following mass
eigenstate phases
\begin{eqnarray*}
\Phi_{\1} & = & E_{\1} \left( T - t_{\1,\c} \right) - p_{\1}
v_{\1} \left( T - t_{\1,\c} \right)~,\\
\Phi_{\2} & = & E_{\2} \left( T - t_{\2,\c} \right) - p_{\2}
v_{\2} \left( T - t_{\2,\c} \right)~.
\end{eqnarray*}
The last two terms in the difference of phase
\begin{equation}
\label{intlg}
\Delta \Phi  = T \, \left[ \Delta E - \Delta (pv) \right] -
\Delta \left(Et_{\c} \right) + \Delta \left( pvt_{\c} \right)
\end{equation}
represent the generalization of Eq.\,(\ref{intl}) in the case of
non instantaneous creation. However, as explained in the previous
Section, in the wave packet formalism additional initial phases
are automatically included in  the expression of the oscillation
phase. Thus, for non instantaneous creation, we still have to take
into account the contributions of the initial phases
\begin{eqnarray*}
\Phi_{\1,\in} & = & E_{\1} t_{\1,\c} - p_{\1}
\left[ x_{\fp} - v_{\1} \left( T - t_{\1,\c} \right) \right]~,\\
\Phi_{\2,\in} & = & E_{\2} t_{\2,\c} - p_{\2}
\left[ x_{\fp} - v_{\2} \left( T - t_{\2,\c} \right) \right]~.
\end{eqnarray*}
The final result contains both the difference of phase (\ref{intlg})
and the additional term
\begin{equation}
\label{intlgi}
\Delta \Phi_{\in}  = - \Delta p \, x_{\fp} +T \,  \Delta (pv) +
\Delta \left(Et_{\c} \right) - \Delta \left( pvt_{\c} \right)~.
\end{equation}
Finally, after integration ($x_{\fp} \to \bar{L}$)
the standard result
\begin{equation}
\Delta \Phi_{\st} =T \, \Delta E - \bar{L} \,  \Delta p
\end{equation}
is once more obtained.

The above procedure has eliminated flavor contamination at
creation. Nevertheless, since creation no longer occurs at a
unique time and the partially formed wave packets naturally
evolve, there will still not be a pure flavor eigenstate at any
fixed time, with the trivial exception of the very first instant
in the creation process. We also observe that any search for the
particle during this time (creation) will not necessarily yield a
positive result since the wave function is not fully normalized.
The measurement will still produce a collapse of the wave function
in the appropriate percentage of cases to zero.

\section*{VI. GEDANKEN WAVE PACKETS}

We have seen that the implementation  of pure flavor creation,
while non trivial, does not modify the standard oscillation
formula. However, the wave packet assumed  was by no means the
most general. Now we shall study the possible consequences of a
two-peak wave packet. This is the simplest which allows for the
insertion of additional constant initial phase factors.

To simplify the following calculation, we return to the unphysical
(in any frame) $\Delta p = 0$ scenario with instantaneous creation
at $t=0$. We consider a wave packet obtained by a sum of
generalized Gaussians with peaks at $x = \pm x_{\0}$ (see Fig.\,3),
\begin{equation}
\label{dps} \psi(x,0) = N \, \left\{ \exp \left[ i \, p_{\0} (x +
x_{\0}) \right]  \, \exp \left[ - \frac{\, \, (x +
x_{\0})^{\2}}{\, \, a^{\2}} \right]  + \exp \left[ i \, p_{\0} (x-
x_{\0}) \right] \, \exp \left[ - \frac{\, \, (x - x_{\0})^{\2}}{\,
\, a^{\2}} \right] \right\}~,
\end{equation}
where $N$ is the normalization constant. We also assume that the
peaks are well separated. In this case
\[ N \approx \left( \frac{1}{\, 2 \pi a^{\2} } \right)^{\f14}~.\]
Note that each gaussian has its own extra constant phase factor.

We now allow the wave packets to evolve and consider the situation
(measurement) in which the first peak of one mass
eigenstate overlaps with the second peak of the other (see Fig.\,4).
In this region of overlap,  oscillation occurs but it is to be noted that
there are non-overlapping parts of the wave packet which
correspond to pure mass eigenvalues (see Fig.\,5). The contribution of these
parts yields a non oscillating term. Indeed this is exactly what
happens when the mass wave packets have completely separated
(decoherence). The oscillating term is  modified by the presence
of the difference between the phase factors $\exp [ip_{\0}
x_{\0}]$ and $\exp [-i p_{\0} x_{\0}]$. Thus,
the oscillation
formula at or around these times  reads
\begin{equation}
\label{tps}
 P \left( \boldsymbol{\nu_{\a}}  \to \boldsymbol{\nu_{\b}} ; \, t
\right) \approx \sin^{\2} 2 \theta  \left\{  1 -
\mbox{$\frac{1}{2}$} \,
\exp \left[ - \,  \left( \, \frac{\Delta v
\, t - 2 x_{\0} }{a \sqrt{2}} \right)^{\2} \right] \,
\cos \left[ \left( \frac{\Delta m^{\2}}{2
\bar{E}} \, t \right) \mp 2 p_{\0} x_{\0} \right] \right\} \, / \,
2~,
\end{equation}
where the sign in the constant term $\mp \, 2 p_{\0} x_{\0}$  depends
on $v_{\1} \mbox{\footnotesize $\gtrless$} \, v_{\2}$.
In this picture,
\[ 2 p_{\0} x_{\0}
\approx \pm \, p_{\0}  \Delta v \, t \approx \mp \, \frac{\Delta
m^{\2}}{2 \bar{E}} \, t ~.
\]
Thus, we obtain the extra factor two of Section II for the
oscillating phase. It is true that since $\mp \, 2 p_{\0} x_{\0}$
is a constant term the oscillation period is still be the standard
one (see Fig.\,5). However, with multiple peaks, each with
independent initial constant phases, the effective extra phase
term could become $L$ or $T$ dependent. We see no good physical
reasons for excluding such a contribution a priori.

In the case of {\em minimal slippage} when the interference occurs
within each separate peak, the constant phases play no role. In
these cases, we can also ignore (by definition) the
non-overlapping parts of the wave function. Thus, the standard
result is obtained as a good approximation. This is now consistent
with the fact that for no slippage ($\Delta v=0$) the standard
formula is {\em exact}.

It is also useful to observe that, in the incoherent limit, one
has a clean method for determining experimentally the mixing
angle, whatever the wave packet form is,
\[
P \left( \boldsymbol{\nu_{\a}}  \to \boldsymbol{\nu_{\b}} ; \,
T \gg a/ |\Delta v| \right) = \sin^{\2} 2 \theta \, / \, 2~.
\]
Nevertheless, the time for the onset of complete incoherence
depends upon the wave packet dimensions and upon the different
mass velocities.

\section*{VII. CONCLUSIONS}

The aim of this work was to shed some new light on the quantum
mechanics of particle oscillations. The primary objective was to
search for the conditions under which the standard oscillation
formula is valid. In the process we have understood the origin of
the extra factors in the  plane wave oscillation calculations. It
is the implicit assumption that at creation (whether instantaneous
or not) the flavor eigenstate is  unique even up to the phase at
all points and times of creation. To the best of our knowledge
this has not been pointed out previously. The often quoted {\em
plane wave derivations}  of the standard formula have generally
been based upon invalid approximations or formalisms chosen so as
to compensate for the neglect of the initial phase contributions.

Other authors have previously pointed out that the assumptions of
equal velocity, momentum and energy are ``unphysical" in the sense
that they are not compatible with the known production processes
in the laboratory frame. We have pointed out here that the first
two are rigourously {\em non physical} in the sense that there is
no Lorentz frame in which they occur. Only the equal energy case
is theoretically possible. Consequently, since the assumption of
{\em instantaneous creation} together with flavor eigenstate
production imposes equal momentum, it is non physical. We can
correct for this by assuming an {\em event-wise production}
mechanism. Event-wise production is perfectly natural. It is even
predicted for the equal momentum case (had it existed) when seen
by another observer in a Lorentz frame in which the momenta are
not equal. Somewhat surprisingly the standard oscillation phase,
as calculated in the wave packet formalism, is  not affected by
this modification, at least not for the cases with minimal
slippage. Thus, we have concluded that within the wave packet
formalism the standard result is not only exact in the case of
equal velocities (no slippage) but also a good approximation in
all cases in which minimal slippage occurs between the mass wave
packets.

Now the standard oscillation formula contains no dependence upon
the form of the wave packets involved. {\em This clearly cannot be
valid in general}. Indeed, as a simple and well known counter
example we have recalled in the previous Section the incoherence
limit, who's onset is dependent upon the size of the wave packets
and the differences in the mass eigenstate velocities. One must
expect the dimensions and explicit form of the wave packets,
including all relative phases, to play a role in the oscillation
formula. We have exhibited a simple two-gaussian model to
demonstrate not only this but also how a result that simulates the
extra-factors calculations (for the oscillating part) can be
obtained.

Allowing for completely arbitrary modulations of the plane waves,
we must introduce additional parameters into our generalized
formula
\begin{equation}
P \left( \boldsymbol{\nu_{\a}}  \to \boldsymbol{\nu_{\b}} ; \, t
\right) = \sin^{\2} 2 \theta \left\{ 1 - R(t) \, \cos \left[
\frac{\Delta m^{\2}}{2 \bar{E}} \, t + \varphi(t) \right] \right\}
\, / \, 2~,
\end{equation}
where $R(t)$ and $\varphi(t)$  depend upon the details of the wave
packet envelope. This formula contains two extra parameters when
compared to the standard formula, four in all. It may well prove
necessary to employ our generalized expression in order to
reconcile diverse experimental results. The alternative could lead
to inconsistencies in the determination of mass differences and/or
mixing angles. Use of this formula may also avoid the need to
introduce one or more sterile neutrinos. This equation, as it
stands, is probably too vague for practical phenomenological fits.
However, it is always possible, without returning to the standard
formula, to add simplifying assumptions such as the time
independence of the extra phase term.

Apart from the question posed in this paper of what is the most
practical oscillation formula to be used in phenomenological fits,
there is an aspect of this work which is of great interest, at
least to the authors. The details of the creation and annihilation
of wave packets is to a large degree unknown territory. Oscillation
phenomena may indeed be useful as a source of information upon
this subject. It should also be possible to investigate this
aspect for photon wave packets with the help of very precise
measurements in interference phenomena. In the case of photons the
effect of slippage is substituted by the occurrence of different
path lengths. For example, in optics it is well known that
interference effects cease if the difference of path lengths
exceeds the wave packet dimension, and this should permit the
determination of these dimensions. This technique may, of course,
also be extended to any elementary particle that lives long
enough. This particular subject matter recalls transitory
phenomena in various sectors of classical physics. Its study, both
theoretical and experimental, at the quantum mechanical level is
surely a great challenge.

\newpage

\begin{figure}[hbp]
\includegraphics[width=10cm, height=17cm, angle=-90]{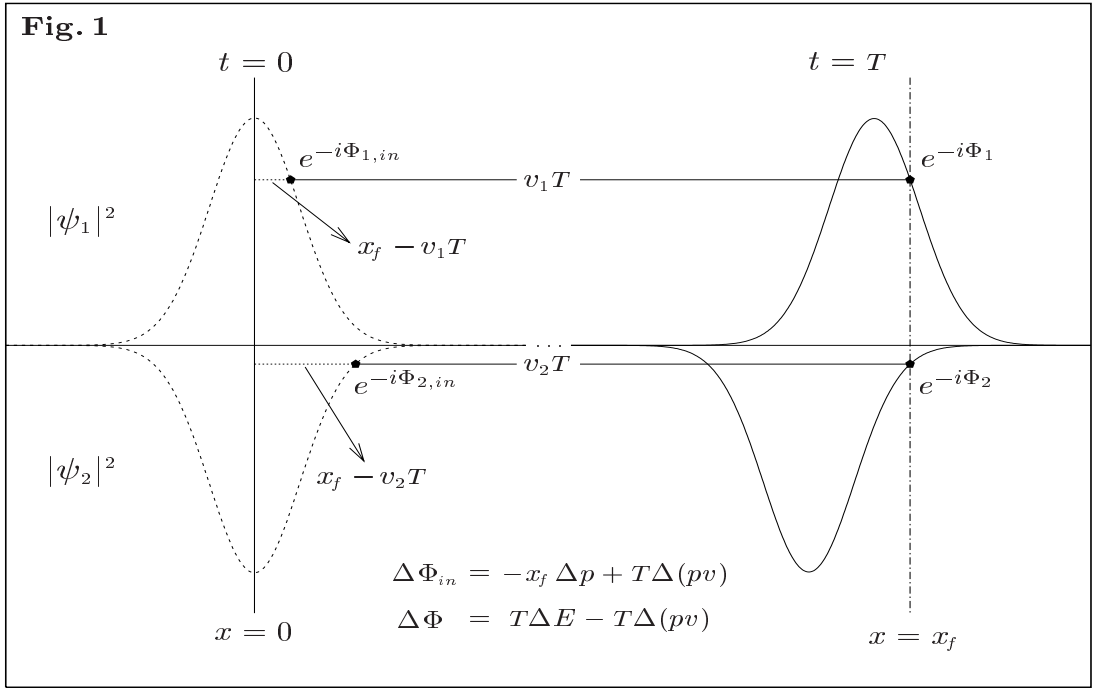}
\caption{ The square modulus of the mass eigenstate coefficients,
$|\psi_{\1}(x,t)|^{\2}$ (upper-half) and $|\psi_{\2}(x,t)|^{\2}$
(lower-half), is plotted as a fucntion of $x$ for two  times:
$t=0$ (left-side) and $t=T$ (right-side). As a model assumption,
the wave packets are supposed to be  Gaussian functions (with the
same width $a$) peaked around $x=0$. At observation time $T$, the
mass eigenstate wave packets are centered around {\em different}
space-points, $L_{\1}=v_{\1}T$ and $L_{\2}=v_{\2}T$ (spreading
effects are neglected). Since $L_{\1}$ and $L_{\2}$ are assumed
very large compared to $a$, the plots of the mass eigenstate wave
packets at creation and observation are separated by dots in the
$x$-axis. At observation time $T$, $x_{\fp}$ is a fixed point in
the overllaping region. At this point, the mass eigenstate phases
are given by $\Phi_{\1}(x_{\fp},T)=E_{\1}T - p_{\1} v_{\1} T$ and
$\Phi_{\2}(x_{\fp},T)=E_{\2}T - p_{\2} v_{\2} T$. The interfering
wave packet points at $(x_{\fp},T)$ correspond to {\em different}
initial wave packet points. Thus, the initial phases
$\Phi_{\1,\in}(x_{\fp}-v_{\1}T,0)= - p_{\1} x_{\fp} + p_{\1}
v_{\1} T$ and $\Phi_{\2,\in}(x_{\fp}-v_{\2}T,0)= - p_{\2} x_{\fp}
+ p_{\2} v_{\2} T$ are automatically included in the wave packet
formalism. Consequently, the {\em standard} result $T \Delta E -
x_{\fp} \Delta p$ is obtained. }
\end{figure}

\newpage

\begin{figure}[hbp]
\includegraphics[width=10cm, height=17cm, angle=-90]{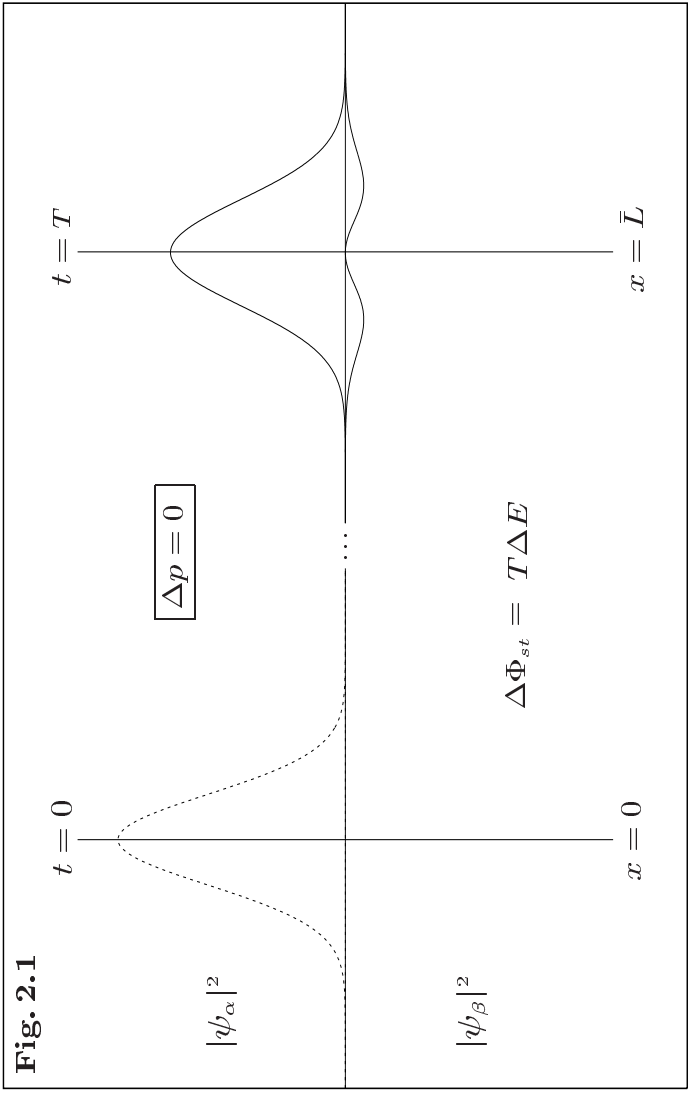}
\includegraphics[width=10cm, height=17cm, angle=-90]{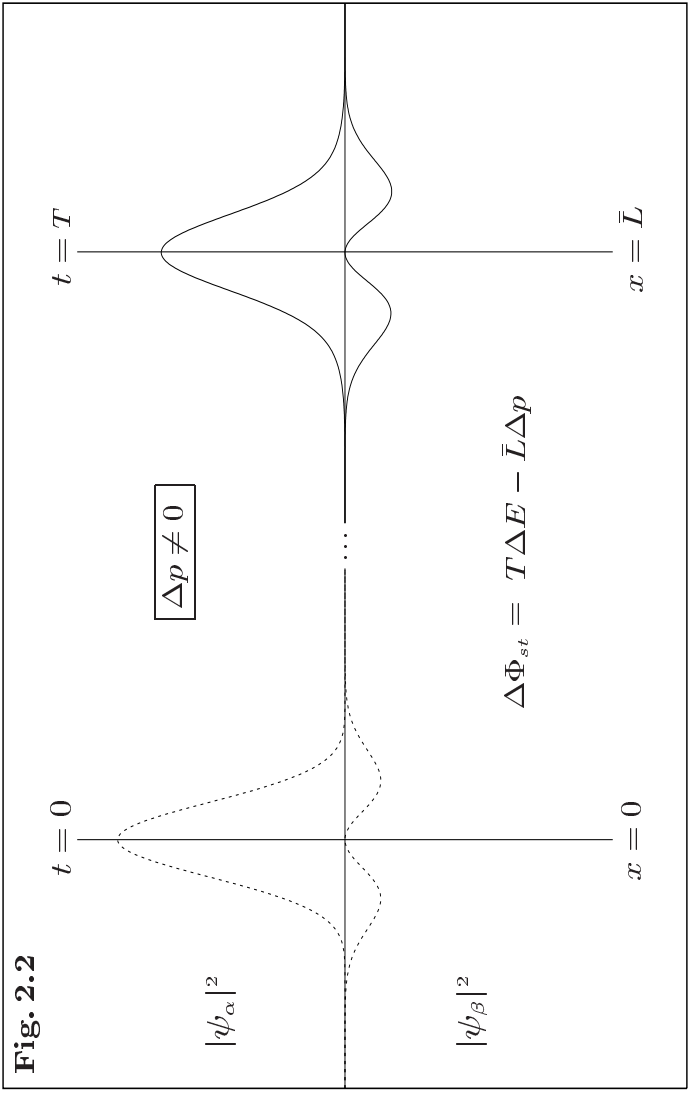}
\caption{ The square modulus of the flavor eigenstate
coefficients, $|\psi_{\a}(x,t;\frac{\pi}{4})|^{\2}$ (upper-half)
and $|\psi_{\b}(x,t;\frac{\pi}{4})|^{\2}$ (lower-half), is plotted
as a function of $x$ for two times: $t=0$ (left-side) and $t=T$
(right-side). The observation time $T$ was chosen as an integer
number of standard oscillation periods, $T_{\osc} =  4 \pi
\bar{E}/ \Delta m^{\2}$. In the case $\Delta p=0$, a pure falvor
state $\boldsymbol{\nu_{\a}}$ is created at time $t=0$ (upper plot).
Slippage
(see Fig.\,1) leads to a non zero probability to find a flavor
state $\boldsymbol{\nu_{\b}}$ at time $T$. However, in the case
$\Delta p\neq 0$, within the context of instantaneous creation,
there does not exist {\em any} time for which the state is a pure
flavor eigenstate (lower plot)}.
\end{figure}

\newpage

\begin{figure}[hbp]
\includegraphics[width=10cm, height=17cm, angle=-90]{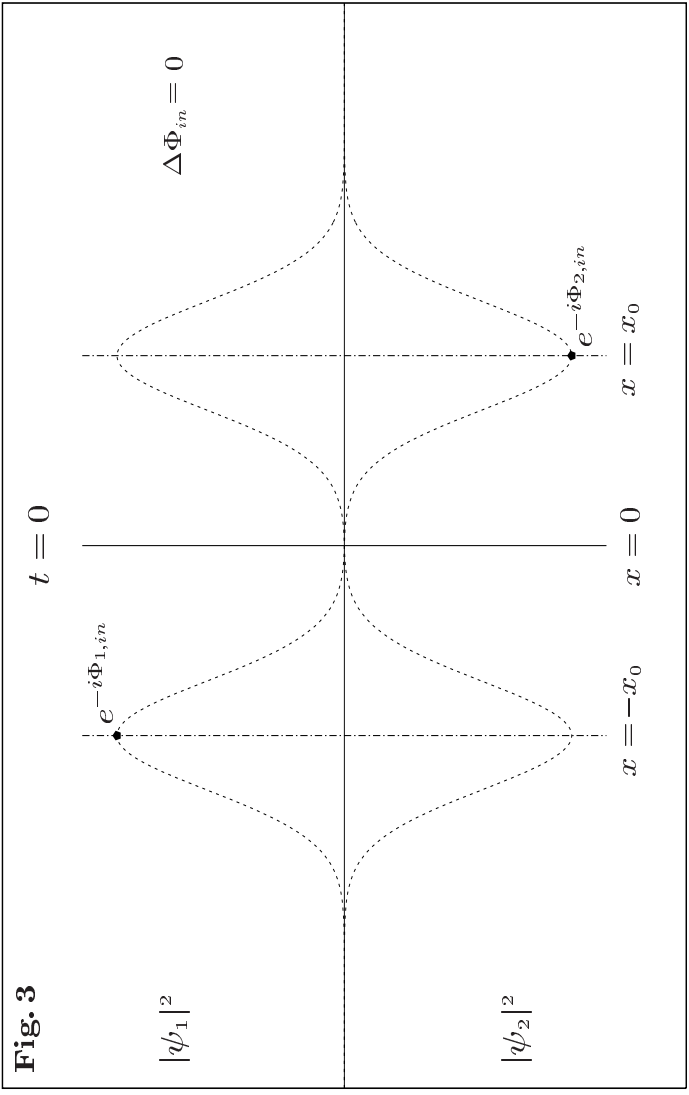}
\caption{ The square modulus of the mass eigenstate coefficients,
$|\psi_{\1}(x,t)|^{\2}$ (upper-half) and $|\psi_{\2}(x,t)|^{\2}$
(lower-half), is plotted as a function of $x$ for the initial time
$t=0$. As a model assumption, the wave packets are supposed to be
{\em generalized} Gaussian  functions peaked around $x=\pm
x_{\0}$.}
\end{figure}

\newpage

\begin{figure}[hbp]
\includegraphics[width=10cm, height=17cm, angle=-90]{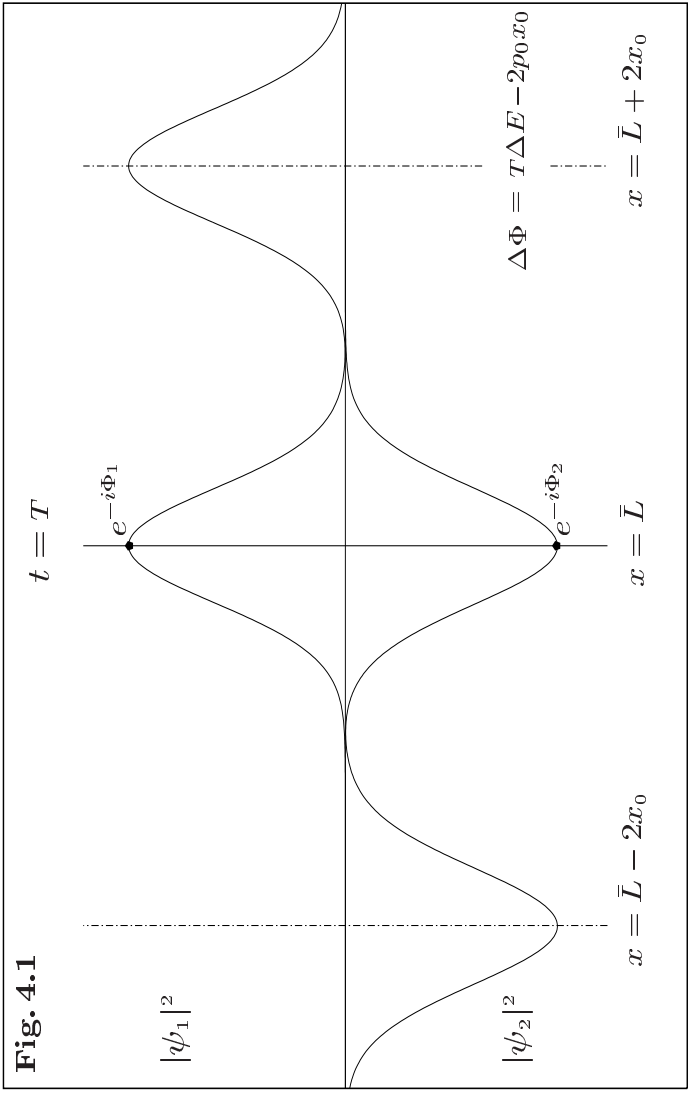}
\includegraphics[width=10cm, height=17cm, angle=-90]{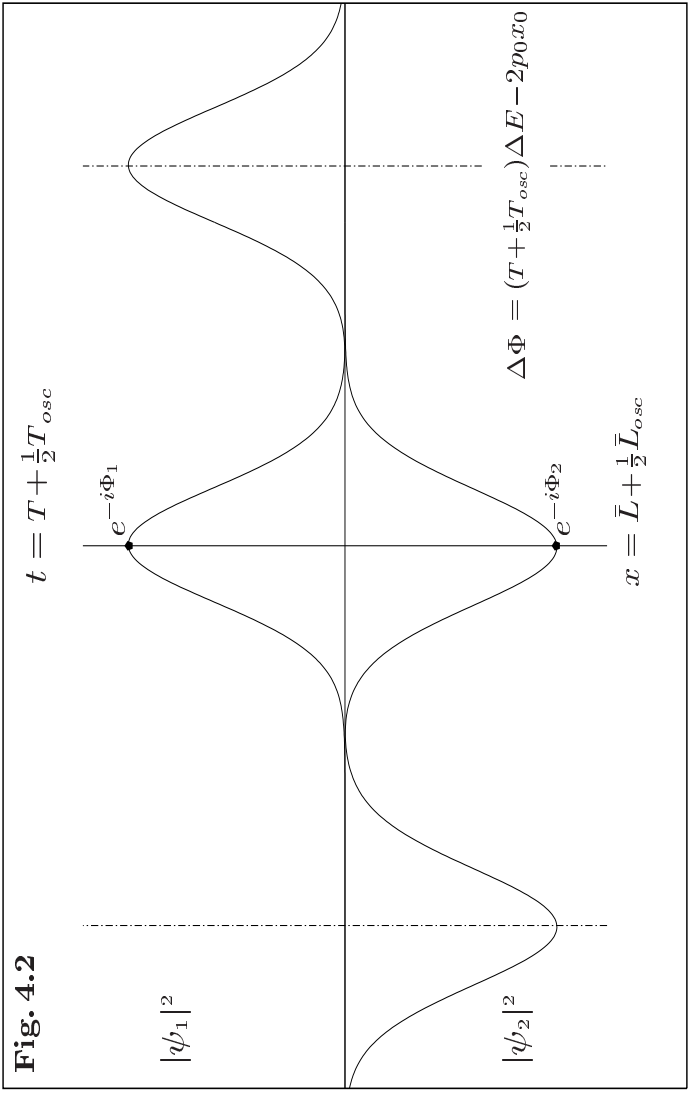}
\caption{ The square modulus of the mass eigenstate coefficients,
$|\psi_{\1}(x,t)|^{\2}$ (upper-half) and $|\psi_{\2}(x,t)|^{\2}$
(lower-half), is plotted as a function of $x$ for two times: $t=T$
and $t=T+\frac{1}{2}\, T_{\osc}$. The choice of $T=2x_{\0} /
\Delta v$ guarantees that the first peak (from the left) of the
$\boldsymbol{\nu_{\1}}$  mass eigenstate wave packet overlaps with
the second  peak of the
 $\boldsymbol{\nu_{\2}}$ mass eigenstate wave packet.
By observing that  $T/ T_{\osc} \approx x_{\0} \bar{E} \gg 1$ and
using Eq.\,(\ref{tps}), it can be immediately understood why the
plots of the mass eigenstate wave packets at the times $t=T$ and
$t=T+\frac{1}{2}\, T_{\osc}$ practically coincide. }
\end{figure}

\newpage

\begin{figure}[hbp]
\includegraphics[width=10cm, height=17cm, angle=-90]{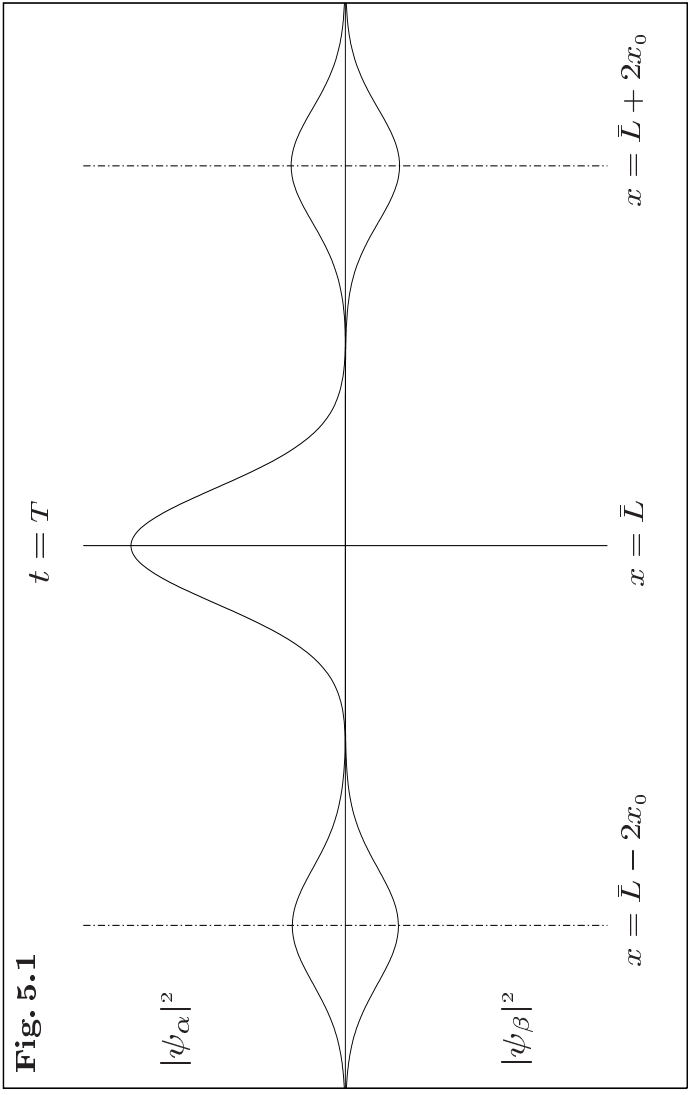}
\includegraphics[width=10cm, height=17cm, angle=-90]{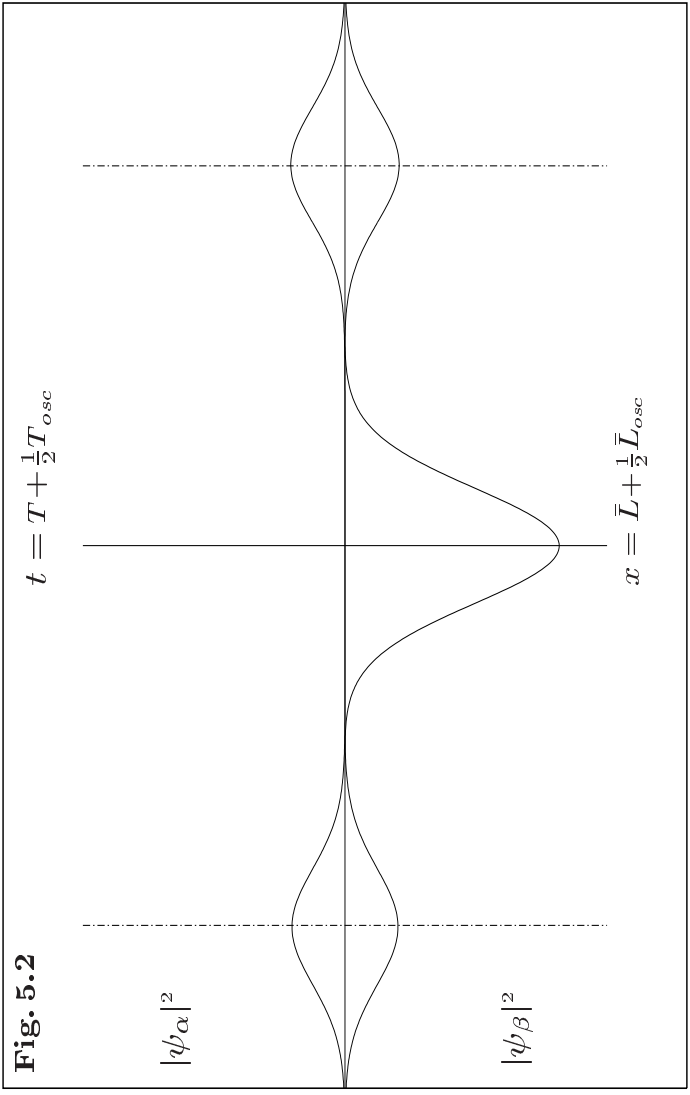}
\caption{ The square modulus of the flavor eigenstate
coefficients, $|\psi_{\a}(x,t;\frac{\pi}{4})|^{\2}$ (upper-half)
and $|\psi_{\b}(x,t;\frac{\pi}{4})|^{\2}$ (lower-half), is plotted
as a function of $x$ for two  times: $t=T$  and $t=T+\frac{1}{2}\,
T_{\osc}$. The choice of $T=2x_{\0} / \Delta v$ (which does {\em
not} necessarily coincides with an integer number of standard
oscillation period $T_{\osc}$) corresponds to a (local) maximum
probability to find the flavor state $\boldsymbol{\nu_{\a}}$
(upper plot). This
would not agree with the prediction of the standard formula,
because of the presence of the extra constant phase term.
Nevertheless, the standard oscillation period is maintained.
Indeed, at the later time $T+\frac{1}{2}\, T_{\osc}$, we have
a (local) maximum probability to find the flavor state
$\boldsymbol{\nu_{\b}}$ (lower plot).}
\end{figure}

\end{document}